\documentclass[%
 reprint,
 amsmath,amssymb,
 aps,
 pra
]{revtex4-2}

\usepackage{graphicx}% Include figure files
\usepackage{dcolumn}% Align table columns on decimal point
\usepackage{bm}% bold math
\usepackage{xcolor}

\usepackage[
  pdfauthor={},
  pdfkeywords={},
  bookmarks=true, linktocpage, colorlinks=true, allbordercolors=white,
allcolors=blue]{hyperref}
\usepackage{orcidlink} % must come AFTER hyperref

\newcommand{\newc}{\newcommand*}

\long\def\begincomment#1\endcomment{%
        \begingroup\sf\baselineskip12pt#1\endgroup}

\newc{\etal}{\textrm{et al.}} 
\newc{\eg}{\textrm{e.g.}} 
\newc{\ie}{\textrm{i.e.}}
\newc{\etc}{\textrm{etc}}
\newc\vs{\textrm{vs.}}
\newc{\cl}{\rm {C.L.}}
\newc{\ev}{\ensuremath{\,\mathrm{eV}}}
\newc{\kev}{\ensuremath{\,\mathrm{keV}}}
\newc{\mev}{\ensuremath{\,\mathrm{MeV}}}
\newc{\gev}{\ensuremath{\,\mathrm{GeV}}}
\newc{\tev}{\ensuremath{\,\mathrm{TeV}}}
\newc{\MeV}{\mev} 
\newc{\TeV}{\tev}
\newc{\invpb}{\ensuremath{/\text{pb}}}
\newc{\invfb}{\ensuremath{/\text{fb}}}
\newc\nb{\ensuremath{\,\mathrm{nb}}} \newc\pb{\ensuremath{\,\mathrm{pb}}} \newc\fb{\ensuremath{\,\mathrm{fb}}}
\newc\pc{\ensuremath{\,\mathrm{pc}}}
\newc\kpc{\ensuremath{\,\mathrm{kpc}}}
\newc\mpc{\ensuremath{\,\mathrm{Mpc}}}
\newc\ps{\ensuremath{\,\mathrm{ps}}} 
\newc\cmeter{\ensuremath{\,\mathrm{cm}}} 
\newc\meter{\ensuremath{\,\mathrm{m}}} 
\newc\kmeter{\ensuremath{\,\mathrm{km}}}
\newc\second{\ensuremath{\,\mathrm{s}}}
\newc\msecond{\ensuremath{\,\mathrm{ms}}}
\newc\nsecond{\ensuremath{\,\mathrm{ns}}}
\newc\psecond{\ensuremath{\,\mathrm{ps}}}
\newc{\chisqmin}{\ensuremath{\chi^2_{\mathrm{min}}}}
\newc{\Delchisq}{\ensuremath{\Delta\chi^2}}
\newc{\chisq}{\ensuremath{\chi^2}}
\newc{\like}{\ensuremath{\mathcal{L}}}
\newc\lsim{\ensuremath{\mathrel{\rlap{\lower4pt\hbox{\hskip1pt$\sim$}}\raise1pt\hbox{$<$}}}}
\newc\gsim{\ensuremath{\mathrel{\rlap{\lower4pt\hbox{\hskip1pt$\sim$}}\raise1pt\hbox{$>$}}}}
\newc{\VEV}[1]{\ensuremath{\langle #1 \rangle}}
\newc{\dl}{\ensuremath{\stackrel{\leftarrow}{D}}}
\newc{\dr}{\ensuremath{\stackrel{\rightarrow}{D}}}
\newc{\bcenter}{\begin{center}}   
\newc{\ecenter}{\end{center}}
\newc{\bfl}{\begin{flushleft}}    
\newc{\efl}{\end{flushleft}}
\newc{\bfr}{\begin{flushright}}   
\newc{\efr}{\end{flushright}}

\newc{\bi}{\begin{itemize}}
\newc{\ei}{\end{itemize}}
\newc{\bed}{\begin{description}}
\newc{\eed}{\end{description}}
\newc{\ben}{\begin{enumerate}}
\newc{\een}{\end{enumerate}}

\newc{\be}{\begin{equation}}
\newc{\ee}{\end{equation}}
\newc{\bea}{\begin{eqnarray}}
\newc{\eea}{\end{eqnarray}}
\newc{\bfle}{\begin{flalign}}
\newc{\efle}{\end{flalign}}
\newc{\ra}{\rightarrow}
\newc{\alphas}{\ensuremath{\alpha_s}}
\newc{\alphatwo}{\ensuremath{\alpha_2}}
\newc{\alphaone}{\ensuremath{\alpha_1}}
\newc{\alphai}[1]{\ensuremath{\alpha_{#1}}}
\newc{\alphaem}{\ensuremath{\alpha_{\mathrm{em}}}}
\newc{\alphaeff}{\ensuremath{\alpha_{\mathrm{eff}}}}
\newc{\sineff}{\ensuremath{\sin \theta_{\mathrm{eff}}}}
\newc{\sinsqeff}{\ensuremath{\sin^2 \theta_{\mathrm{eff}}}}
\newc{\dalphahad}{\ensuremath{\Delta \alpha_{\mathrm{had}}}}
\newc{\yt}{\ensuremath{h_t}} \newc{\yb}{\ensuremath{h_b}} \newc{\ytau}{\ensuremath{h_{\tau}}}
\newc\mz{\ensuremath{M_Z}} 
\newc\mw{\ensuremath{m_W}}
\newc\mZ{\mz}        \newc\mW{\mw}
\newc\mhsm{\ensuremath{ m_{H_{\mathrm{SM}}}}}
\newc{\mtop}{\ensuremath{ m_t}}               \newc{\mtpole}{\ensuremath{ M_t}}
\newc{\mbottom}{\ensuremath{ m_b}} 
\newc{\mtau}{\ensuremath{ m_{\tau}}}
\newc{\mt}{\mtpole}
\newc{\mb}{\mbottom} 
\newc{\rtwogg}{\ensuremath{R_{h_2}(\gamma\gamma)}}
\newc{\rtwozz}{\ensuremath{R_{h_2}(ZZ)}}
\newc{\ronegg}{\ensuremath{R_{h_1}(\gamma\gamma)}}
\newc{\ronezz}{\ensuremath{R_{h_1}(ZZ)}}
\newc{\rsiggg}{\ensuremath{R_{h_\textrm{sig}}(\gamma\gamma)}}
\newc{\rsigzz}{\ensuremath{R_{h_\textrm{sig}}(ZZ)}}
\newc{\llbar}{\ensuremath{\ell\bar{\ell}}}
\newc{\tauptaum}{\ensuremath{ \tau^+\tau^-}}
\newc{\qqbar}{\ensuremath{ q\bar{q}}} \newc{\ppbar}{\ensuremath{ p\bar{p}}}
\newc{\bbbar}{\ensuremath{ b\bar{b}}} \newc{\ttbar}{\ensuremath{ t\bar{t}}}
\newc{\ffbar}{\ensuremath{ f\bar{f}}} \newc{\tautaubar}{\ensuremath{ \tau\bar{\tau}}}

\newc{\mchi}{\ensuremath{m_\neutone}}
\newc{\squark}{\ensuremath{\tilde{q}}}
\newc{\slepton}{\ensuremath{\tilde{l}}}
\newc{\gluino}{\ensuremath{\tilde{g}}} 
\newc{\mgluino}{\ensuremath{{m_{\gluino}}}}
\newc{\wino}{\ensuremath{\tilde{W}}} 
\newc{\mwino}{\ensuremath{{m_{\wino}}}}
\newc{\tone}{\ensuremath{{\tilde{t}_1}}}
\newc{\Hone}{\ensuremath{{\tilde{H}_{1}}}}
\newc{\Htwo}{\ensuremath{{\tilde{H}_{2}}}}
\newc{\Hhtwo}{\ensuremath{{H_{2}}}}
\newc{\qli}{\ensuremath{{\tilde{Q}_{i}}}}
\newc{\uri}{\ensuremath{{\tilde{u}_{i}}}}
\newc{\dri}{\ensuremath{{\tilde{d}_{i}}}}
\newc{\lli}{\ensuremath{{\tilde{L}_{i}}}}
\newc{\eri}{\ensuremath{{\tilde{e}_{i}}}}
\newc{\sthw}{\ensuremath{ \sin\theta_W}}              \newc{\cthw}{\ensuremath{\cos\theta_W}}
\newc{\tanthw}{\ensuremath{ \tan\theta_W}}              \newc{\cotthw}{\ensuremath{\cot\theta_W}}
\newc{\ssqthw}{\ensuremath{\sin^2 \theta_W}}
\newc{\msbar}{\ensuremath{\overline{MS}}} \newc{\drbar}{\ensuremath{\overline{DR}}}
\newc{\mtmtsmmsbar}{\ensuremath{ m_t(m_t)^{\msbar}_{{\mathrm{SM}}}}}
\newc{\mtmtsmdrbar}{\ensuremath{ m_t(m_t)^{\drbar}_{{\mathrm{SM}}}}}
\newc{\mtmtmssmdrbar}{\ensuremath{ m_t(m_t)^{\drbar}_{{\mathrm{SUSY}}}}}
\newc{\mbmbmsbar}{\ensuremath{ m_b(m_b)^{\msbar} }}
\newc{\mbmbsmmsbar}{\ensuremath{ m_b(m_b)^{\msbar}_{{\mathrm{SM}}}}}
\newc{\mbmzsmmsbar}{\ensuremath{ m_b(\mz)^{\msbar}_{{\mathrm{SM}}}}}
\newc{\mbmzsmdrbar}{\ensuremath{ m_b(\mz)^{\drbar}_{{\mathrm{SM}}}}}
\newc{\mbmzmssmdrbar}{\ensuremath{ m_b(\mz)^{\drbar}_{{\mathrm{SUSY}}}}}
\newc{\mtaumzsmmsbar}{\ensuremath{ m_{\tau}(\mz)^{\msbar}_{{\mathrm{SM}}}}}
\newc{\mtaumzsmdrbar}{\ensuremath{ m_{\tau}(\mz)^{\drbar}_{{\mathrm{SM}}}}}
\newc{\mtaumzmssmdrbar}{\ensuremath{ m_{\tau}(\mz)^{\drbar}_{{\mathrm{SUSY}}}}}
\newc{\alphasmzms}{\ensuremath{\alpha_s(M_Z)^{\overline{MS}}}}
\newc{\alphaimzms}[1]{\ensuremath{\alpha_{#1}(M_Z)^{\overline{MS}}}}
\newc{\alphaemmz}{\ensuremath{\alpha_{\mathrm{em}}(M_Z)^{\overline{MS}}}}
\newc{\mzero}{\ensuremath{{m_0}}}
\newc{\mhalf}{\ensuremath{ m_{1/2}}}
\newc{\tanb}{\ensuremath{\tan\beta}}
\newc{\azero}{\ensuremath{ A_0}}
\newc{\signmu}{\ensuremath{\rm{sgn}\,\mu}}
\newc{\atau}{\ensuremath{{A_{\tau}}}}
\newc{\mueff}{\ensuremath{\mu_{\rm{eff}}}}
\newc{\lam}{\ensuremath{{\lambda}}}
\newc{\kap}{\ensuremath{{\kappa}}}
\newc{\alam}{\ensuremath{{A_{\lambda}}}}
\newc{\akap}{\ensuremath{{A_{\kappa}}}}
\newc{\hs}{\ensuremath{ H_s}}      
\newc{\mhs}{\ensuremath{ m_{H_s}}} 
\newc{\mgut}{\ensuremath{ M_{\rm GUT}}}
\newc{\mvl}{\ensuremath{ M_{\rm VL}}}
\newc{\gut}{\ensuremath{{\rm GUT}}}
\newc{\mplanck}{\ensuremath{ M_{\rm P}}}      \newc{\mpl}{\ensuremath{ M_{\rm Pl}}}
\newc{\msusy}{\ensuremath{ M_{\rm SUSY}}}      \newc{\ms}{\ensuremath{ M_{\rm S}}}
 \newc{\hu}{\ensuremath{ H_u}}       \newc{\hd}{\ensuremath{ H_d}}
 \newc{\mhu}{\ensuremath{ m_{H_u}}}       \newc{\mhd}{\ensuremath{ m_{H_d}}}
 \newc{\mhuew}{\ensuremath{ m^{\ast}_{H_u}}}       \newc{\mhdew}{\ensuremath{ m^{\ast}_{H_d}}}
 \newc{\mhuewsq}{\ensuremath{ m^{\ast\, 2}_{H_u}}}       \newc{\mhdewsq}{\ensuremath{ m^{\ast\, 2}_{H_d}}}
 \newc{\mhl}{\ensuremath{m_\hl}} 
 \newc{\mhone}{\ensuremath{m_{h_1}}} 
 \newc{\mhtwo}{\ensuremath{m_{h_2}}} 
 \newc{\mhi}{\ensuremath{m_{\tilde{h}}}} 
 \newc{\mul}{\ensuremath{m_{\tilde{u}_L}}} 
 \newc{\mtone}{\ensuremath{m_{\tilde{t}_1}}} 
 \newc{\ma}{\ensuremath{m_A}} 
 \newc{\mH}{\ensuremath{m_H}} 
 \newc{\maone}{\ensuremath{m_{a_1}}} 
 \newc{\matwo}{\ensuremath{m_{a_2}}}
 \newc{\hone}{\ensuremath{h_1}}
 \newc{\htwo}{\ensuremath{h_2}}
 \newc{\aone}{\ensuremath{a_1}}
 \newc{\atwo}{\ensuremath{a_2}}
 \newc{\mqthree}{\ensuremath{m_{\tilde{Q}_3}^2}}
 \newc{\muthree}{\ensuremath{m_{\tilde{u}_3}^2}}
 \newc{\mqli}{\ensuremath{m_{\tilde{Q}_{i}}}}
 \newc{\muri}{\ensuremath{m_{\tilde{u}_{i}}}}
 \newc{\mdri}{\ensuremath{m_{\tilde{d}_{i}}}}
 \newc{\mlli}{\ensuremath{m_{\tilde{L}_{i}}}}
 \newc{\meri}{\ensuremath{m_{\tilde{e}_{i}}}}
 \newc{\ts}{\ensuremath{T_{SUSY}}}

\newc{\sigsip}{\ensuremath{\sigma^{\rm SI}_{p}}}	\newc{\sigsin}{\ensuremath{\sigma^{\rm SI}_{n}}}
\newc{\sigsdp}{\ensuremath{\sigma^{\rm SD}_{p}}}	\newc{\sigsdn}{\ensuremath{\sigma^{\rm SD}_{n}}}
\newc{\sigsi}{\ensuremath{\sigma^{\rm SI}}}	\newc{\sigsd}{\ensuremath{\sigma^{\rm SD}}}
\newc{\abund}{\ensuremath{ \Omega h^2}}
\newc{\omegadm}{\ensuremath{ \Omega_{{\rm DM}}}}     \newc{\abunddm}{\ensuremath{ \Omega_{{\rm DM}} h^2}} 
\newc{\omegam}{\ensuremath{ \Omega_{{\rm m}}}}       \newc{\abundm}{\ensuremath{ \Omega_{{\rm m}} h^2}}
\newc{\omegab}{\ensuremath{ \Omega_{{\rm b}}}}	\newc{\abundb}{\ensuremath{ \Omega_{{\rm b}} h^2}}
\newc{\omegatot}{\ensuremath{ \Omega_{{\rm TOT}}}}
\newc{\omegacdm}{\ensuremath{ \Omega_{{\rm CDM}}}}   \newc{\abundcdm}{\ensuremath{ \Omega_{{\rm CDM}} h^2}}
\newc{\omegalambda}{\ensuremath{ \Omega_{\Lambda}}} \newc{\abundlambda}{\ensuremath{ \Omega_{\Lambda} h^2}}
\newc{\omegarad}{\ensuremath{ \Omega_{{\rm rad}}}}  \newc{\abundrad}{\ensuremath{ \Omega_{{\rm rad}} h^2}}
\newc{\rhocrit}{\ensuremath{ \rho_{\rm crit}}}
\newc{\rhochi}{\ensuremath{ \rho_{\chi}}}
\newc{\abunchi}{\ensuremath{\Omega_\chi h^2}}
\newc{\abundlsp}{\ensuremath{\Omega_{\rm LSP}h^2}}
\newc{\amu}{\ensuremath{ a_{\mu}}}        \newc{\amususy}{\ensuremath{ a_{\mu}^{\mathrm{SUSY}}}}
\newc{\amuexpt}{\ensuremath{ a_{\mu}^{\mathrm{expt}}}}        \newc{\amusm}{\ensuremath{ a_{\mu}^{\mathrm{SM}}}}
\newc\deltaamu{\ensuremath{\Delta a_{\mu}}} \newc{\deltaamususy}{\ensuremath{\delta a_{\mu}^{\mathrm{SUSY}}}}
\newc\gmtwo{\ensuremath{ (g-2)_{\mu}}} 
\newc{\deltagmtwomususy}{\ensuremath{\delta\left(g-2\right)_{\mu}^{\mathrm{SUSY}}}}
\newc{\deltagmtwomu}{\ensuremath{\delta\left(g-2\right)_{\mu}}}
\newc{\deltagmtwoe}{\ensuremath{\delta\left(g-2\right)_{e}}}
\newc\BR{\ensuremath{\rm BR}}
\newc\bsgamma{\ensuremath{ b\rightarrow s \gamma }}
\newc\bxsgamma{\ensuremath{\overline{B}\rightarrow X_{s}\gamma}}
\newc\brbsgamma{\ensuremath{\BR\left(\bsgamma\right)}}
\newc\brbxsgamma{\ensuremath{\BR\left(\bxsgamma\right)}}
\newc\bsmumu{\ensuremath{B_s\to\mu^+\mu^-}}
\newc\brbsmumu{\ensuremath{\BR\left(B_s\to\mu^+\mu^-\right)}}
\newc\bdmmumu{\ensuremath{\overline{B}_d\to\mu^+\mu^-}}
\newc\bbbarmix{\ensuremath{\overline{B}_s\mbox{-}B_s}}      % B_s mixing
\newc\delmbs{\ensuremath{\Delta M_{B_s}}}
\newc{\butaunu}{\ensuremath{B_u \rightarrow \tau \nu}}
\newc{\brbutaunu}{\ensuremath{\BR\left(B_u \rightarrow \tau \nu\right)}}
\newcommand*{\reftable}[1]{Table~\ref{#1}}         
       
\newcommand*{\reffig}[1]{Fig.~\ref{#1}}
 
        \newcommand*{\refeq}[1]{Eq.~(\ref{#1})}
     \newcommand*{\refsec}[1]{Sec.~\ref{#1}}

\newcommand*{\neutone}{\ensuremath{\tilde{\chi}^0_1}}

\let\oldcite\cite
\renewcommand*{\cite}{~\oldcite}

\newcommand*{\hl}{\ensuremath{h}}

\begin{document}

\preprint{APS/123-QED}

\title{GUT-induced collider signatures of the LUX-ZEPLIN event}% Force line breaks with \\
%\thanks{A footnote to the article title}%

\author{Wojciech Kotlarski\orcidlink{0000-0002-1191-6343}}
 \email{wojciech.kotlarski@ncbj.gov.pl}
\author{Kamila Kowalska\orcidlink{0000-0001-8457-2847}}
 \email{kamila.kowalska@ncbj.gov.pl}
 \author{Enrico Maria Sessolo\orcidlink{0000-0001-6571-6382}}%
 \email{enrico.sessolo@ncbj.gov.pl}
\affiliation{%
 National Centre for Nuclear Research, Pasteura 7, 02-093 Warsaw, Poland }%

\date{\today}% It is always \today, today,
             %  but any date may be explicitly specified
             
%%%%%%%%%%%%%%%%%%%%%%%%%%%%%%%%%%%%%%%%%%%%%
\begin{abstract}
The recent observation of an event with large recoil energy at LUX-ZEPLIN~(LZ) has found a possible explanation in terms of the inelastic scattering of a pseudo-Dirac particle, for example, the Higgsino of the MSSM. For pseudo-Dirac dark matter,
the mass splitting $\delta m$ is correlated with the mass of some heavier states. While in the MSSM rough dimensional considerations seem to indicate mass ranges of these heavy particles hopelessly out of reach even at future colliders, we point out that this is not the case in GUT-induced scenarios, with or without supersymmetry. Working in a non-supersymmetric model based on SU(6), we show that the experimentally favored $\delta m$ implies the presence of a $Z'$ gauge boson with a mass of several~TeVs. We derive the predicted $Z'$ signal in the dilepton channel and we confront it with the sensitivity of the HL-LHC and the FCC-$hh$ to $Z'$ resonances. We show that the $Z'$ falls squarely in reach of the discovery potential of both experiments.
\end{abstract}

\maketitle

%\tableofcontents
%%%%%%%%%%%%%%%%%%%%%%%%%%%%%%%%%%%%%%%%%%%%%
\section{Introduction}

Recently, the LUX-ZEPLIN~(LZ) Collaboration reported the results of a search for weakly interacting massive particles~(WIMPs) in xenon, extending the nuclear recoil energy range up to approximately 270\kev\cite{LZ:2026axp}. The analysis identified an event of interest, consistent with a nuclear recoil energy of $248\pm 23(\textrm{stat})\pm 23(\textrm{sys})\kev$. The profile likelihood test favors 
the dark matter~(DM) hypothesis over the background only, globally by $2.6\sigma$, with a maximum local significance of $3.4\sigma$. The interpretation strongly points to
the inelastic scattering of DM rather than the standard elastic spin-independent scattering, due to the incompatible shape of the energy spectrum in the standard case.

The LZ excess has generated excitement in the community, with a substantial amount of speculation about its origin\cite{Su:2026rwz,Fan:2026kxx,Freese:2026sga,Wu:2026nhi,Lou:2026idn,Yin:2026jnn,DiMauro:2026ldr,Pospelov:2026ewn,Visinelli:2026kgt,Yamashita:2026ump,Chattopadhyay:2026ryw,Smirnov:2026aqk,Du:2026guj,Rodd:2026tyn,McCabe:2026crm,Jeesun:2026vzo,Unwin:2026rdp}. 
One of the most exciting possibilities\cite{Fan:2026kxx,Freese:2026sga,Wu:2026nhi,Yin:2026jnn,DiMauro:2026ldr,Pospelov:2026ewn,Du:2026guj,Rodd:2026tyn} is that the recoil is due to the inelastic scattering of a Higgsino, the SU(2)-doublet Majorana fermion of the minimal supersymmetric standard model~(MSSM), which is one of the very few WIMP candidates remaining unscathed by ever more stringent experimental constraints\cite{Krall:2017xij,Roszkowski:2017nbc,Kowalska:2018toh}. In order to generate a tree-level 
splitting  of the right order of magnitude in the neutral Higgsino states, the gauginos of the MSSM are expected to be quite heavy: $\delta m\approx m_Z^2/M_{1,2}\approx 250\kev$ implies 
$M_{1,2}\approx 10^7\gev$, hopelessly out of reach in any future collider search. 

Interestingly, the LZ signal may find an 
explanation in a WIMP scenario that features more promising detection prospects than in the MSSM. 
In fact, it is well known that any combination of Majorana-Dirac multiplets can potentially lead to an inelastic scattering signal like the one potentially observed by LZ. Such combinations include  
doublet-singlet mixing\cite{DiMauro:2026ldr} but also the mixing of other representations\cite{Smirnov:2026aqk}. In turn, this calls for the analysis of more complex models, for example scenarios with UV boundary 
conditions rooted in a Grand Unified Theory~(GUT). 

Incidentally, GUTs provide an elegant framework to accommodate the desired particle content beyond the Standard Model~(BSM). But more importantly, the breaking chain of a GUT down to the Standard Model~(SM) group often leads to
the appearance of an extra gauge symmetry, typically a U(1)$_X$, which has to be spontaneously broken at some intermediate scale below $M_{\textrm{GUT}}\approx 10^{16}\gev$. 
In this paper, we will show that the $Z'$ gauge boson associated with the breaking of U(1)$_X$ can acquire a mass directly related to the mass splitting of a neutral pseudo-Dirac DM candidate with characteristics very similar to those of a Higgsino. Unlike in the MSSM case, however, the $Z'$ particle of GUT constructions falls squarely within the early reach of a future hadron collider. In this paper, we work out the sensitivity of the high-luminosity~(HL-)LHC and the FCC in the clean dilepton signal channel.

Note that most non-supersymmetric GUT completions do not usually lead to a fermionic DM candidate. 
The breaking chain $\textrm{SO(10)}\to\textrm{SU(5)}\times\textrm{U(1)}\to\textrm{SM}$
does potentially contain fermion DM\cite{Ferrari:2018rey}, since the particles of the dark and visible sectors are separated by their U(1) charges. 
This, however, is typically not the case with GUTs based on $\textrm{SU($N$)}\to\textrm{SU(5)}\times\textrm{SU($N-5$)}\to\textrm{SM}$\cite{Rizzo:2022lpm}. In those cases, DM is either a pseudo-Nambu-Goldstone boson\cite{Cacciapaglia:2019ixa,Cai:2020njb,Otsuka:2022zdy,Chiang:2023omu}, or one is forced to introduce extra symmetries, as was done, e.g., in Ref.\cite{Ma:2020hyy,Okada:2018tgy}. Note that the fermion DM emerging from an added $\mathbb{Z}_2$ symmetry is typically a singlet of SU(2)$_L$\cite{Ma:2020hyy,Okada:2018tgy}. 
On the other hand, some of us have shown in a recent study\cite{Chikkaballi:2025pnw}, 
that more options for DM, 
including a mixture of SU(2) doublets and singlets affine to the Higgsino, become available if 
the UV dynamics that prevents the DM particle from decaying is rooted in 
\textit{quantum scale symmetry}, i.e., the presence of irrelevant Gaussian fixed points in the 
renormalization group flow of the Yukawa couplings. 
In such a scenario, the coupling strengths of the gauge-Yukawa system are entirely predicted by the UV completion. 

The paper is organized as follows. In \refsec{sec:signal} we recall the kinematics of the DM signal observed by LZ and the favored range of mass splitting for inelastic DM. In \refsec{sec:model} we review the non-supersymmetric SU(6) GUT model introduced in Ref.\cite{Chikkaballi:2025pnw} and the UV mechanism that prevents the DM from decaying. We give parametric expressions for the pseudo-Dirac DM mass and splitting and relate them to the mass of a $Z'$ gauge boson originating from the breaking of the GUT symmetry. We dedicate \refsec{sec:dm_solar} to discussing the impact of solar-capture rate bounds in our model and the ways to avoid them.
In \refsec{sec:collider}, we perform a collider analysis of the discovery reach of the HL-LHC at $\sqrt{s}=14\tev$ and of the FCC-$hh$ at $\sqrt{s}=100\tev$ for resonances in the dilepton channel. We show that the $Z'$ mass correlated to the DM mass splitting in SU(6) falls squarely within the projected sensitivity of both experiments. We summarize our findings and conclude in \refsec{sec:conc}. Appendix~\ref{app:scap} is dedicated to a brief review of the low-scale scalar potential of the model. 

%%%%%%%%%%%%%%%%%%%%%%%%%%%%%%%%%%%%%%%%%%%%%%%%
\section{The DM signal\label{sec:signal}}
%%%%%%%%%%%%%%%%%%%%%%%%%%%%%%%%%%%%%%%%%%%%%%%%

At direct detection experiments, the differential nuclear recoil rate as a function of recoil energy $E_R$ reads
\be
\frac{dR}{d E_R}\approx \frac{\sigma_n}{2\, m_{\textrm{DM}} \mu_n^2}\left(A-Z \right)^2 F^2(E_R) 
\mathcal{G}\left(v_{\textrm{min}}(E_R,\delta m),v_{\textrm{esc}}\right),
\ee
where $\sigma_n$ is the DM-nucleon scattering cross section, $m_{\textrm{DM}}$ is the DM mass, $\mu_n$ is the reduced mass of the DM-nucleon system, $A-Z$ is the number of neutrons in the nuclear target, $F(E_R)$ is the nuclear form factor of the target nucleus\cite{Vietze:2014vsa}, and $\mathcal{G}$ parametrizes the DM velocity in the halo. 

Typically, a Maxwell-Boltzmann distribution $f(\mathbf{v},v_0)$ is used so that
\be
\mathcal{G}\left(v_{\textrm{min}}(E_R,\delta m),v_{\textrm{esc}}\right)
=\rho_0\int_{v_{\textrm{min}}<|\mathbf{v}|<v_{\textrm{esc}}}\frac{f(\mathbf{v},v_0)}{|\mathbf{v}|}d^3 v,
\ee
which depends on the local DM density, $\rho_0\approx 0.4\gev/\textrm{cm}^3$, the circular velocity of the DM halo, $v_0\approx 220\,\textrm{km}/\textrm{s}$, and the Galactic escape velocity, $v_{\textrm{esc}}\approx 540\,\textrm{km}/\textrm{s}$. 
Another possibility\cite{Besla:2019xbx,Smith-Orlik:2023kyl}, explored in the context of the LZ~excess in Refs.\cite{Fan:2026kxx,Wu:2026nhi,Freese:2026mfi}, is that the Standard Halo Model~(SHM) velocity distribution is enhanced by interactions with neighboring halos, such as the Large Magellanic Cloud~(LMC). Such interactions can accelerate the DM and significantly contribute to the high-speed tail of the local velocity distribution.

In the case of inelastic scattering, minimal velocity $v_{\textrm{min}}$ depends explicitly on the mass splitting, $\delta m$, between the initial and final state
\cite{Peter:2013aha,Fan:2026kxx}:
\be
v_{\textrm{min}}(E_R,\delta m)=\sqrt{\frac{m_N E_R}{2 \mu_N^2}}+\frac{\delta m}{\sqrt{2 E_R m_N}}\,,
\ee
where $m_N\approx 122\gev$ is the xenon nucleus mass, and $\mu_N$ is the reduced mass of the DM-nucleus system. As was pointed out in Refs.\cite{Fan:2026kxx,Bramante:2016rdh}, when $m_{\textrm{DM}}\gg \mu_N$, one finds that $v_{\textrm{min}}$ is minimized at $E_R\approx \delta m$. Thus, the LZ determination leads directly to mass splittings in the ballpark of~$250\kev$ for a relatively heavy WIMP DM candidate. Recent fits based on the SHM\cite{Freese:2026sga,Fan:2026kxx} point to $\delta m\approx 340-360\kev$ for $m_{\textrm{DM}}=1.1\tev$. Including the LMC, the fits\cite{Fan:2026kxx,Wu:2026nhi,Freese:2026mfi} show preference for a larger mass splitting, $\delta m\approx 475-520\kev$ at $m_{\textrm{DM}}=1.1\tev$. 

%%%%%%%%%%%%%%%%%%%%%%%%%%%%%%%%%%%%%%%%%%%%%%%%
\section{The model\label{sec:model}}
%%%%%%%%%%%%%%%%%%%%%%%%%%%%%%%%%%%%%%%%%%%%%%%%

Consider a GUT based on SU($N$) with $N\geq 6$, populated with a certain number of fermion~$(F)$ and scalar~$(S)$ multiplets. In order to ensure anomaly cancellation, one necessitates some specific fermion content. One minimal solution employs $N-4$ copies of the antifundamental, $\overline{\Box}^{(F)}_{\textrm{SU($N$)}}$, and one antisymmetric representation, $\textbf{anti}^{(F)}_{\textrm{SU($N$)}}$, per fermion generation. 

Consider next the GUT-breaking chain $\textrm{SU($N$)}\to\textrm{SU(5)}\times\textrm{SU($N-5$)}\to\textrm{SM}$. One finds in this case,
\be
\begin{aligned}
\overline{\Box}^{(F)}_{\textrm{SU($N$)}}&\to (\mathbf{\bar{5}},\mathbf{1})^{(F)}+(\mathbf{1},\overline{\Box}_{\textrm{SU($N-5$)}})^{(F)}\\
\textbf{anti}^{(F)}_{\textrm{SU($N$)}}&\to (\mathbf{10},\mathbf{1})^{(F)}+(\mathbf{5},\Box_{\textrm{SU($N-5$)}})^{(F)}+\, ...
\end{aligned}
\ee

In order to break SU($N$) spontaneously down to SU(5) and also guarantee a nonzero mass for all light massive particles, a certain number of scalar multiplets will be needed. Typically, this will involve the adjoint of SU($N$) and several other scalar representations matching the SU($N$) fermion representations derived from anomaly cancellation. The Yukawa sector of the SM belongs to the well-known SU(5) Lagrangian terms
\be
Y_U \mathbf{10}^{(F)} \mathbf{10}^{(F)} \mathbf{5}^{(S)}+Y_D \mathbf{10}^{(F)} \mathbf{\bar{5}}^{(F)} \mathbf{\bar{5}}^{(S)}+\textrm{H.c.}\,,
\ee
where, in the SU($N$) embedding, $\mathbf{10}^{(F)}\subset \textbf{anti}^{(F)}_{\textrm{SU($N$)}}$ 
and  $ \mathbf{\bar{5}}^{(F)}\subset\overline{\Box}^{(F)}_{\textrm{SU($N$)}}$\,. 
Note that one can always arrange to isolate at least two, eventually light, 
color-neutral BSM doublets of SU(2) in the SU($N$) system: 
one sitting in a copy of $\overline{\Box}^{(F)}_{\textrm{SU($N$)}}$ different from the one carrying the SM, 
and the other sitting in one of the $(\mathbf{5},\Box_{\textrm{SU($N-5$)}})^{(F)}$ of $\textbf{anti}^{(F)}_{\textrm{SU($N$)}}$. One can also find at least two light BSM singlets of the SM group in the $N-4$ copies of $(\mathbf{1},\overline{\Box}_{\textrm{SU($N-5$)}})^{(F)}$. The mixing of color-neutral doublets and singlets, which is then generated when the scalar sector acquires vacuum expectation values~(vevs), provides the ingredients we need for a non-supersymmetric DM~candidate with characteristics similar to the MSSM Higgsino. Moreover, the same sector typically contains enough ingredients to generate the neutrino masses via the seesaw mechanism\cite{Ma:2020hyy}.

Since the breaking of SU($N$) leaves the scalar fields that give mass to the DM charged under SU($N-5$), new gauge bosons will appear in the low energy theory. Their mass will be related to the inelastic mass splitting $\delta m\approx y_{\textrm{lep}}^2 v_{\textrm{SM}}^2/(y_{\textrm{BSM}}\,v_{\textrm{BSM}})$,  
similarly to the way gaugino masses are related to $\delta m$ in the MSSM. However, because the DM carries lepton number, Yukawa coupling $y_{\textrm{lep}}$ is expected to be much smaller than the SM gauge couplings. Consequently, given equivalent $\delta m$, we expect $v_{\textrm{BSM}}$ to be correspondingly smaller than the gaugino masses, so much so to be possibly in reach of planned future hadron colliders.

Let us work out the formalism in a concrete example based on SU(6). The fermionic content, determined by anomaly cancellation, includes three Weyl multiplets per generation: 
$\mathbf{\bar{6}_1}^{(F)}$, $\mathbf{\bar{6}_2}^{(F)}$, and $\mathbf{15}^{(F)}$. To make sure that all fermions acquire masses after the gauge symmetries are spontaneously broken,  
at least three scalar multiplets need to be introduced. Following Ref.\cite{Chikkaballi:2025pnw}, 
we are going to consider 
four scalar multiplets, $\mathbf{6_1}^{(S)}$, $\mathbf{6_2}^{(S)}$, $\mathbf{15}^{(S)}$, and $\mathbf{21}^{(S)}$.
One also needs the adjoint
$\mathbf{35}^{(S)}$, which breaks $\textrm{SU(6)}\to\textrm{SU(5)}\times\textrm{U(1)}_X$.

Let us now define the Yukawa couplings consistent with SU(6) symmetry. For each fermion generation, the Lagrangian reads,
\be\label{eq:yuk}
\begin{aligned}
\mathcal{L} \supset{} & 
y_u\, \mathbf{15}^{(F)} \mathbf{15}^{(F)} \mathbf{15}^{(S)}+
y_{11} \mathbf{15}^{(F)} \mathbf{\bar{6}_1}^{(F)} \mathbf{\bar{6}_1}^{(S)}\\
& + y_{12} \mathbf{15}^{(F)} \mathbf{\bar{6}_1}^{(F)} \mathbf{\bar{6}_2}^{(S)}+ y_{21} \mathbf{15}^{(F)} \mathbf{\bar{6}_2}^{(F)} \mathbf{\bar{6}_1}^{(S)}  \\
& +y_{22} \mathbf{15}^{(F)} \mathbf{\bar{6}_2}^{(F)} \mathbf{\bar{6}_2}^{(S)} +\,\tilde{y}_{11}\, \mathbf{\bar{6}_1}^{(F)} \mathbf{\bar{6}_1}^{(F)} \mathbf{15}^{(S)}\\
&+\tilde{y}_{12}\, \mathbf{\bar{6}_1}^{(F)} \mathbf{\bar{6}_2}^{(F)} \mathbf{15}^{(S)} 
 +\tilde{y}_{22}\, \mathbf{\bar{6}_2}^{(F)} \mathbf{\bar{6}_2}^{(F)} \mathbf{15}^{(S)}\\
 &+\,\hat{y}_{11}\, \mathbf{\bar{6}_1}^{(F)} \mathbf{\bar{6}_1}^{(F)} \mathbf{21}^{(S)} + \hat{y}_{12}\, \mathbf{\bar{6}_1}^{(F)} \mathbf{\bar{6}_2}^{(F)} \mathbf{21}^{(S)}  \\
&+\hat{y}_{22}\, \mathbf{\bar{6}_2}^{(F)}  \mathbf{\bar{6}_2}^{(F)} \mathbf{21}^{(S)}+\textrm{H.c.} 
\end{aligned}
\ee
We will carry out 
our analysis for the third generation only. In other words, we are neglecting flavor-mixing effects in the first approximation.

The scalar potential of the model was given in Ref.\cite{Chikkaballi:2025pnw} and we recall it in Appendix~\ref{app:scap}. Its free parameters will have an impact on the discussion in \refsec{sec:dm_solar}.

We adopt the following SU(6)-breaking chain:
\begin{itemize}
\item $v_6=\langle \mathbf{35}^{(S)}\rangle\approx 10^{16}\gev$ breaks $\textrm{SU(6)}\to\textrm{SU(5)}\times\textrm{U(1)}_X$
\item $v_5=\langle \mathbf{24}_0^{(S)}\rangle\approx 10^{16}\gev$ breaks $\textrm{SU(5)}\to \textrm{SU(3)}_\textrm{c}\times \textrm{SU(2)}_L\times\textrm{U(1)}_Y$
\item $\mathbf{\bar{6}_1}^{(S)}\supset\mathbf{\bar{5}}^{(S)}_{-1}\supset\left(\mathbf{1},\mathbf{\bar{2}},-\frac{1}{2};-1\right)+\left(\mathbf{\bar{3}},\mathbf{1},\frac{1}{3};-1\right)$, where we have indicated
in parentheses the quantum numbers in the extended SM group, $\textrm{SU(3)}_\textrm{c}\times \textrm{SU(2)}_L\times\textrm{U(1)}_Y\times \textrm{U(1)}_X$. 
Doublet-triplet splitting follows from the tuning of some scalar potential couplings (see Appendix~A.1 in Ref.\cite{Chikkaballi:2025pnw}) and it assures that the SU(2) doublet field remains light
\item $\mathbf{\bar{6}_2}^{(S)}\supset\mathbf{1}^{(S)}_{5}=\left(\mathbf{1},\mathbf{1},0;5\right)$ is a SM-singlet scalar field that remains light
\item $\mathbf{15}^{(S)}\supset\mathbf{5}^{(S)}_{-4}\supset\left(\mathbf{1},\mathbf{2},\frac{1}{2};-4\right)+\left(\mathbf{3},\mathbf{1},-\frac{1}{3};-4\right)$. Doublet-triplet splitting is employed to make the SU(2) doublet field remain light
\item $\mathbf{21}^{(S)}\supset\mathbf{1}^{(S)}_{-10}=\left(\mathbf{1},\mathbf{1},0; -10\right)$ is a SM-singlet scalar that remains light.
\end{itemize}

At collider energies, 
one is left with two SU(2)-doublet scalars (Higgs doublets) and two complex scalar SM-singlets. We rename the light scalar fields for simplicity,
\bea\label{eq:lightf}
H_d=\left(\mathbf{1},\mathbf{\bar{2}},-\frac{1}{2};-1\right)\,, & \quad & H_u=\left(\mathbf{1},\mathbf{2},\frac{1}{2};-4\right)\,,\nonumber\\
s_6=\left(\mathbf{1},\mathbf{1},0;5\right)\,, & \quad & s_{21}=\left(\mathbf{1},\mathbf{1},0;-10\right)\,.
\eea
Throughout the paper, we shall denote weak isospin doublets with capital letters, and weak singlets in lowercase. 
Since no scalar field is neutral under the U(1)$_X$ group, 
the heaviest of the acquired vevs will approximately determine the
mass of the $Z'$ gauge boson, modulo the size of gauge coupling $g_X$ and U(1)$_X$ charge of the scalars. The vevs of the light scalar fields, $v_d$, $v_u$, $v_{s_6}$, $v_{s_{21}}$, give mass to the light fermions in the model. As is customary, we define $\tanb\equiv v_u/v_d$. 

We can define the low-scale, left-chiral Weyl fermion multiplets of the extended SM group
$\textrm{SU(3)}_\textrm{c}\times \textrm{SU(2)}_L\times\textrm{U(1)}_Y\times \textrm{U(1)}_X$. For each generation, the corresponding quantum numbers are summarized in \reftable{tab:q_numbers}.

%%%%%%%%%%%%%%%%%%%%%%%%%%%%%%%%
\begin{table}[t]
\centering
\begin{tabular}{|c|c|c|c|c|c|c|c|c|}
\hline
 & $\;\;Q\;\;$ & $\;u\;$ & $d_{1,2}$ & $\;d'\;$ & $L_{1,2}$ & $\;L'\;$ & $\;\;e\;\;$ & $\nu_{1,2}$ \\
\hline
SU(3)$_\textrm{c}$ & $\mathbf{3}$ & $\mathbf{\bar{3}}$ & $\mathbf{\bar{3}}$ & $\mathbf{3}$ & $\mathbf{1}$ & $\mathbf{1}$ & $\mathbf{1}$ & $\mathbf{1}$ \\
\hline
SU(2)$_L$ & $\mathbf{2}$ & $\mathbf{1}$ & $\mathbf{1}$ & $\mathbf{1}$ & $\mathbf{2}$ & $\mathbf{\bar{2}}$ & $\mathbf{1}$ & $\mathbf{1}$ \\
\hline
U(1)$_Y$ & $\frac{1}{6}$ & $-\frac{2}{3}$ & $\frac{1}{3}$ & $-\frac{1}{3}$ & $-\frac{1}{2}$ & $\frac{1}{2}$ & $1$ & $0$ \\
\hline
U(1)$_X$ & $2$ & $2$ & $-1$ & $-4$ & $-1$ & $-4$ & $2$ & $5$ \\
\hline
\end{tabular}
\caption{Quantum numbers of the low-scale, left-chiral Weyl fermion multiplets in the extended SM gauge group.}
\label{tab:q_numbers}
\end{table}
%%%%%%%%%%%%%%%%%%%%%%%%%%%%%%

For completeness, we indicate the original SU(6) representations from which the fermions emerge upon the breaking of the GUT symmetry:
\be
\begin{aligned}
\mathbf{\bar{6}_1}^{(F)}\supset\mathbf{\bar{5}}^{(\textrm{SM})}_{-1}\supset d_1,\,L_1 & \quad \mathbf{\bar{6}_1}^{(F)}\supset\mathbf{1}^{(F)}_{5}\supset \nu_1 \\
\mathbf{\bar{6}_2}^{(F)}\supset\mathbf{\bar{5}}^{(F)}_{-1}\supset d_2,\,L_2 & \quad \mathbf{\bar{6}_2}^{(F)}\supset\mathbf{1}^{(F)}_{5}\supset \nu_2 \\
\mathbf{15}^{(F)}\supset\mathbf{10}^{(\textrm{SM})}_{2}\supset Q,\, u,\, e &\quad \mathbf{15}^{(F)}\supset\mathbf{5}^{(F)}_{-4}\supset d',\, L'\,.
\end{aligned}
\ee 

We do not concern ourselves in this work with the 
details of gauge-coupling unification. It is well known\cite{Ma:2020hyy} that unification can be achieved at the scale $M_{\textrm{GUT}}\approx 4\times 10^{16}\gev$, if one adds to the low-energy model some extra fields: a color-octet fermion and a weak isospin-triplet fermion, 
which above the GUT scale should belong to an adjoint $\mathbf{35}$.  
We neglect in
\refeq{eq:yuk}
possible Yukawa couplings involving the adjoint needed exclusively to guarantee unification. They are absorbed in the details of the UV completion. 

The low-scale Yukawa Lagrangian of the model reads
\begin{multline}\label{eq:extra_lagr}
\mathcal{L}_{\textrm{full}} \supset  
\mathcal{L}_{\textrm{q.s.s.}}+\left[ y'_d\, d_2 H_d Q +y'_e\, e H_d L_2 + y'_{\nu}\, L' H_d^{c\dag} \nu_2 \right. \\
+\,y'_D\, d_1 d' s_6 + y'_L\, L' L_1 s_6+2\tilde{y}_{11}\,\nu_1 H_u^{c\dag} L_1+2 \tilde{y}_{22}\, \nu_2 H_u^{c\dag} L_2\\
\left. +\,\tilde{y}_{12}\left(\nu_1 H_u^{c\dag} L_2+\nu_2 H_u^{c\dag} L_1 \right)+\hat{y}_{12}\,\nu_1 \nu_2 s_{21}+\textrm{H.c.} \right].
\end{multline}
$\mathcal{L}_{\textrm{q.s.s.}}$, the only part of the Lagrangian that survives 
once quantum scale symmetry is imposed, is given by 
\begin{multline}\label{eq:le_lagr}
\mathcal{L}_{\textrm{q.s.s.}} =  
2 y_u\, u H_u^{c\dag} Q + y_d\, d_1 H_d Q +  y_e\, e H_d L_1 +y_{\nu}\, L' H_d^{c\dag} \nu_1 \\
+y_D\, d_2 d' s_6 +\, y_L\, L' L_2 s_6
+ y_{\nu_1}\, \nu_1 \nu_1 s_{21} + y_{\nu_2}\, \nu_2 \nu_2 s_{21}
+\textrm{H.c.}
\end{multline}
Couplings $y_d$, $y_e$, and $y_{\nu}$ originate from the GUT parameter $y_{11}$ in \refeq{eq:yuk}, 
after following the renormalization group flow down to the electroweak scale.
Couplings $y_D$ and $y_L$ originate from $y_{22}$. 
Couplings $y'_d$, $y'_e$, $y'_{\nu}$ depart from the common value, $y_{21}$, after the breaking of SU(6). Couplings $y'_D$, $y'_L$ stem from~$y_{12}$\,.  

In the basis $\langle d_1, d_2|,\,|d_{Q}, d'\rangle$, and $\langle e, e_{L'}|,\,|e_{L_1}, e_{L_2}\rangle$, 
the mass matrices of the bottom quark and tau lepton read
\begin{equation}\label{eq:VL}
M_b=\frac{1}{\sqrt{2}}\left(\begin{array}{cc}
y_d v_d & y'_D v_{s_6}  \\
y'_d v_d &  y_D v_{s_6}
\end{array}\right),\,
M_{\tau}=\frac{1}{\sqrt{2}}\left(\begin{array}{cc}
y_e v_d & y'_e v_d  \\
y'_L v_{s_6} &  y_L v_{s_6}
\end{array}\right).
\end{equation}

In the basis 
$\langle \nu_{L_1}, \nu_{L_2} ,  \nu_{L'}, \nu_1, \nu_2  |,\, 
|  \nu_{L_1}, \nu_{L_2} ,  \nu_{L'}, \nu_1, \nu_2 \rangle$,
the mass matrix of the neutral Majorana fermions reads
\be\label{eq:fullmass}
 M_{\nu}=\frac{1}{\sqrt{2}}\left(\begin{array}{ccccc}
0 & 0 & y'_L v_{s_6} & 2 \tilde{y}_{11} v_u & \tilde{y}_{12} v_u  \\
0 & 0  & y_L v_{s_6}  & \tilde{y}_{12} v_u & 2 \tilde{y}_{22} v_u\\
y'_L v_{s_6} & y_L v_{s_6}  & 0 & y_{\nu} v_d &  y'_{\nu} v_d \\
2 \tilde{y}_{11} v_u & \tilde{y}_{12} v_u & y_{\nu} v_d   & 2 y_{\nu_1} v_{s_{21}} & \hat{y}_{12} v_{s_{21}}  \\
\tilde{y}_{12} v_u  & 2 \tilde{y}_{22} v_u   & y'_{\nu} v_d   & \hat{y}_{12} v_{s_{21}} & 2 y_{\nu_2} v_{s_{21}} 
\end{array}\right).
\ee

Equation~(\ref{eq:fullmass}) contains several neutral particles that could play the role of a DM WIMP. 
However, its being non-diagonal indicates that no symmetry is in place to protect these heavy particles from mixing with the neutrinos of the SM. Such mixings open up prompt decay channels, e.g., $\textrm{DM}\to h\, \nu_{\textrm{SM}}$, $\textrm{DM}\to Z\, \nu_{\textrm{SM}}$, $\textrm{DM}\to W^{\pm} \tau^{\mp}$, and others. In order to close these dangerous decay channels, one may decide to impose a discrete or a global symmetry, as was done in Refs.\cite{Ma:2020hyy,Okada:2018tgy} 
for SU(6) and SU(5)$\times$U(1), respectively.
We here follow our previous study\cite{Chikkaballi:2025pnw} and impose instead quantum scale symmetry as a symmetry of the trans-Planckian UV.  

%%%%%%%%%%%%%%%%%%%%%%%%%%%%%
\subsection{Vanishing couplings from quantum scale symmetry}

Quantum scale symmetry\cite{Wetterich:1987fm,Shaposhnikov:2008xb,Wetterich:2019qzx,Wetterich:2020cxq} 
is tightly related to asymptotic safety (see Ref.\cite{Eichhorn:2026uqj} and references therein for an up-to-date, modern review), which is the property of a 
quantum field theory to develop
UV fixed points of the renormalization group flow of the quantum effective action\cite{inbookWS}. 

In a theory with fixed points, the coefficients of the operators of the quantum effective action may be divided in two broad classes on the basis of their scaling behavior in the vicinity of the fixed point. Such scaling behavior is determined by the stability matrix, 
\be\label{stab}
M_{ij}=\partial\beta_i/\partial\alpha_j|_{\{\alpha^{\ast}_k\}}\,,
\ee
where $\beta_i$ are the beta functions, $\alpha_j$ are the couplings, 
and the asterisk indicates the fixed-point value. 
Critical exponents $\theta_i$ are defined as the opposite of eigenvalues of the stability matrix and they characterize the power-law evolution of the matter couplings in the vicinity of the fixed point. If $\theta_i<0$, the corresponding eigendirection is dubbed as \textit{irrelevant} or, equivalently, IR-attractive. In this case, only one trajectory connects the UV fixed point to the IR value of the couplings, so that irrelevant couplings either provide specific predictions of phenomenological interest (if the fixed point is nonzero) or, alternatively, are prevented from appearing in the theory at all scales (if the fixed point is zero). In this sense, quantum scale symmetry enhances the symmetry of the gauge theory by forbidding the appearance of certain couplings which would be otherwise allowed.

Conversely, $\theta_i>0$ corresponds to \textit{relevant} eigendirections or, equivalently, UV-attractive. As all renormalization group trajectories along a relevant direction asymptotically reach the fixed point in the deep UV, independently of the measured value of the associated couplings (hence the ``attractiveness''), 
they count as the
free parameters of the macroscopic theory and play the role of effective quantities that cannot be predicted from first principles but can only be determined experimentally. 
Finally, for completeness, we recall that $\theta_i=0$ corresponds to a \textit{marginal} eigendirection. The RG running is logarithmically slow along this direction and an analysis beyond the linear order is required in order to determine whether a fixed point is UV-attractive or IR-attractive.

One important property of the renormalization group flow of irrelevant couplings is that they can be parametrized as unique functions of the relevant couplings' flow. Thus, in a theory with just a finite number of relevant parameters, all one needs to determine the full dynamics of the quantum effective action is the corresponding finite number of experimental measurements. This astonishing quality goes under the name of \textit{non-perturbative renormalizability}. The hypothesis that it is a property of quantum gravity is corroborated by theoretical evidence accumulated over the course of three decades and provides the cornerstone of the field of asymptotically safe quantum gravity\cite{Reuter:1996cp,Codello:2008vh,Reuter:2012id,Pawlowski:2020qer,Eichhorn:2026uqj}.

%%%%%%%%%%%%%%%%%%%%%%%%%%%%%%%%%%%%%%%%%%%%%%%%%%%%%%%%%%%%%%%
\begin{table}[t]
\centering
\begin{tabular}{|c|c|c|c|c|c||c|c|}
\hline
SU(6) & $y_u$ & \multicolumn{2}{c|}{$y_{22}$} & $\hat{y}_{11}$ & $\hat{y}_{22}$ & \multicolumn{2}{c|}{$y_{11}$} \\
\hline
 & $y_u$ & $y_{D}$ & $y_{L}$ & $y_{\nu_1}$ & $y_{\nu_2}$ & $y_{d}$ & $y_{\nu}$ \\
\hline
$\mu=1\tev$ & $0.69$ & $1.1$ & $0.55$ & $0.57$ & $0.51$ & $0.027$ & $0.014$ \\
\hline
\end{tabular}
\caption{(Left-hand side) The predicted values of the irrelevant Yukawa couplings at the scale $\mu=1\tev$ for $\tanb=1$. The upper line indicates the corresponding SU(6) Yukawa couplings.
(Right-hand side) Relevant couplings $y_d$ and $y_{\nu}$ 
are related to each other by the renormalization group flow. Their value at the low scale 
is adjusted so that $y_d v_d/\sqrt{2}$ matches the bottom quark mass\cite{Chikkaballi:2025pnw}.}
\label{tab:le_values}
\end{table}
%%%%%%%%%%%%%%%%%%%%%%%%%%%%%%%%%%%%%%%%%%%%%%%%%%%%%%%%%%%%%%%%%

In the framework of GUT models, UV completions with asymptotic safety where analyzed in Refs.\cite{Eichhorn:2019dhg,Eichhorn:2021qet,Held:2022hnw,Giacometti:2026zrs}. The SU(6) model considered here was investigated in this context by some of us in Ref.\cite{Chikkaballi:2025pnw} and we direct the reader to Appendices~B and C of that work for the renormalization group equations and a set of viable UV fixed points. At the most predictive (irrelevant) fixed point, we found that six Yukawa couplings are prevented from appearing in the Lagrangian by trans-Planckian scale symmetry:
\be\label{eq:fp}
y_{12}=y_{21}=\hat{y}_{12}=\tilde{y}_{12}=\tilde{y}_{11}=\tilde{y}_{22}=0\quad (\textrm{at all scales}).
\ee
The U(1)$_X$ gauge coupling, $g_X$, is predicted by the GUT boundary conditions and reads $g_X=0.07$ at scale~$\mu=1\tev$.
Additionally, four couplings are predicted with specific nonzero values by the trans-Planckian dynamics of the SU(6) system, once a reference value for the top Yukawa coupling is chosen. We show in \reftable{tab:le_values} the values of the nonzero Yukawa couplings at the scale~$\mu=1\tev$ for~$\tanb=1$. It was shown in Ref.\cite{Kotlarski:2023mmr}, that the nonzero values of the gauge and Yukawa couplings predicted from fixed-point boundary conditions are very robust under a broad range of theoretical uncertainties.

%%%%%%%%%%%%%%%%%%%%%%%%%%%%%%%%%%%%%%%%%%%%%%%%
\subsection{Dark matter phenomenology\label{sec:dm_pheno}}
%%%%%%%%%%%%%%%%%%%%%%%%%%%%%%%%%%%%%%%%%%%%%%%%

As some of the Lagrangian couplings allowed by gauge invariance are forbidden by the quantum scale symmetry, low-energy particles that belong to GUT multiplets with the same quantum numbers do not mix with one another. One can thus identify in \refeq{eq:fullmass} a set of four Majorana potential DM candidates: a pair of SU(2) doublets, $N_1$ and $N_2$, and two SU(2) singlets, $N_3$ and $N_4$. In the limit $v_{s_{6}}\ll v_{s_{21}}$ their masses read
\bea
m_{N_{1}}&\simeq& \frac{1}{\sqrt{2}} \left(y_{L}v_{s_{6}}-\frac{1}{4}\frac{y^2_{\nu}v_d^2}{y_{\nu_1}v_{s_{21}}}\right)\,,\nonumber\\
m_{N_{2}}&\simeq& \frac{1}{\sqrt{2}}\left(y_{L}v_{s_{6}}+\frac{1}{4}\frac{y^2_{\nu}v_d^2}{y_{\nu_1}v_{s_{21}}}\right)\,,\nonumber\\ 
m_{N_3}&\simeq & \sqrt{2}\, y_{\nu_2}v_{s_{21}}\,,\nonumber\\
m_{N_4}&\simeq& \sqrt{2} \,y_{\nu_1}v_{s_{21}}\,.\label{eq:physmas}
\eea

It was shown in Ref.\cite{Chikkaballi:2025pnw} that boundary conditions stemming from fixed point~(\ref{eq:fp}) lead to a scenario with two-component DM. It was also shown, however, that selecting a slightly less predictive fixed point can generate a dynamically small\cite{Kowalska:2022ypk,Chikkaballi:2023cce,Eichhorn:2022vgp} coupling $0\neq \tilde{y}_{22}\ll 1$, inducing the decay of one DM candidate onto the other. In the resulting single-candidate scenario, the DM is a pseudo-Dirac doublet of SU(2)$_L$ if $v_{s_6}\ll v_{s_{21}}$. It is a mixture of the neutral components of $L_2,L'$, see~\reftable{tab:q_numbers}. In the early Universe, the two neutral states co-annihilate with their nearly degenerate charged counterparts onto the SM gauge bosons. All four states freeze out, and the relic abundance is saturated at $m_{\textrm{DM}}=m_{N_1}\approx 1.1\tev$, 
in total affinity to the Higgsino case (we neglect possible corrections from the Sommerfeld enhancement for the purposes of this paper).  

%%%%%%%%%%%%%%%%%%%%%%%%%%%%%%%%%%%%%%%%%%%%%%%%%%%%%%%%%%%%%%%%%%%%%%%
 \begin{figure}[t]
	\centering%
		\includegraphics[width=0.4\textwidth]{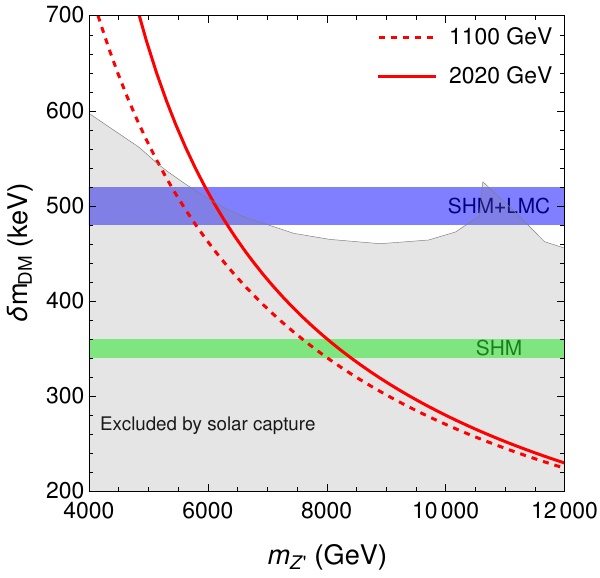}
\caption{Red lines show the dependence of $\delta m_{\textrm{DM}}$ on $m_{Z'}$ in the SU(6) model, given the couplings 
in \reftable{tab:le_values}. In solid, $m_{\textrm{DM}}=2020\gev$; in dashed, $m_{\textrm{DM}}=1100\gev$.
The parameter space favored by LZ is marked in green for the SHM only assumption and in blue for the SHM+LMC assumption. The shaded area shows the recast 90\%~C.L. solar-capture bound on $\delta m_{\textrm{DM}}$ in terms of $m_{Z'}$.}
\label{fig:DD_inelastic}
\end{figure}
%%%%%%%%%%%%%%%%%%%%%%%%%%%%%%%%%%%%%%%%%%%%%%%%%%%%%%%%%%%%%%%%%%%%%

In the limit $v_{s_{6}}\ll v_{s_{21}}$, the mass splitting of the inelastic states reads
\be\label{eq:mass_sp}
\delta m_{\textrm{DM}}=m_{N_2}-m_{N_1}\simeq \frac{1}{2 \sqrt{2}}\frac{y^2_{\nu}v_d^2}{y_{\nu_1}v_{s_{21}}}\,.
\ee

Note that we predict $y_{\nu}=0.014$, see \reftable{tab:le_values}.
As was pointed out in \refsec{sec:model}, since this leptonic Yukawa coupling is much smaller than the SM gauge couplings, we correspondingly expect the denominator in \refeq{eq:mass_sp} to be much smaller than in the Higgsino case.
The mass of the $Z'$ gauge boson is 
\be
\label{eq:mz}
m_{Z'}\simeq 5 g_X \sqrt{v_{s_6}^2+4 v_{s_{21}}^2}\,.
\ee
As it depends on the same parameters as the DM mass and splitting, one can derive the dependence of $\delta m_{\textrm{DM}}$ on $m_{Z'}$. It is shown in \reffig{fig:DD_inelastic}, where the dashed red line corresponds to $m_{\textrm{DM}}=1.1\tev$. The value of $\delta m_{\textrm{DM}}$ favored by the SHM fits\cite{Freese:2026sga,Fan:2026kxx} is shown as a green band, while the value obtained by adding the LMC velocity enhancement\cite{Fan:2026kxx,Wu:2026nhi,Freese:2026mfi} is shown as a blue band.
As is expected, $m_{Z'}< 10^4\gev$ in both cases, lighter by orders of magnitude than the mass of the gauginos predicted by $\delta m_{\textrm{DM}}$ in the MSSM.

%%%%%%%%%%%%%%%%%%%%%%%%%%%%%%%%%%%%%%%%%%%%%%%%
\subsection{Bounds from the solar capture of DM\label{sec:dm_solar}}
%%%%%%%%%%%%%%%%%%%%%%%%%%%%%%%%%%%%%%%%%%%%%%%%

It was first pointed out in Ref.\cite{Pospelov:2026ewn} that, for mass splitting $\delta m_{\textrm{DM}}\lesssim 550\kev$, the thermal Higgsino at 1.1\tev\ 
is in strong tension with the bounds imposed by the non-observation at IceCube of a neutrino flux consistent with the gravitational capture of DM by the Sun\cite{IceCube:2025fcu}. The same conclusion was subsequently reached independently in Refs.\cite{Wu:2026nhi,Nguyen:2026lui,Freese:2026mfi}.

We recast the 90\%~C.L. lower bound on 
$\delta m_{\textrm{DM}}$ from Ref.\cite{Freese:2026mfi} in terms of $m_{Z'}$ in our model. It is shown as a gray exclusion area in \reffig{fig:DD_inelastic}. We have chosen the strongest exclusion in Ref.\cite{Freese:2026mfi} to account for the presence of a nonnegligible spin-dependent cross section for DM-proton elastic scattering, which brings the captured DM to thermal equilibrium inside the Sun. The current bound excludes a~$1.1\tev$ DM mass for both choices of the halo distribution. 

On the other hand, our GUT model contains by construction 
the ingredients allowing us to evade the solar capture bound in the SHM+LMC case. This is possible in two ways. Each solution relies on the standard freeze-out picture and does not alter our prediction of a $Z'$ gauge boson with $m_{Z'}\approx 5.5 - 6\tev$. 
\begin{enumerate}
\item It was shown in Ref.\cite{Freese:2026mfi} that the solar capture bound can be evaded if the SU(2)-doublet WIMP has a mass larger than~$\sim 2\tev$ under the SHM+LMC assumption, and larger than~$\sim 10\tev$ under the SHM only assumption.
As is well known, in the case of a $\sim 2\tev$ SU(2)-doublet Majorana fermion the effective cross section for the coannihilation of its charged and neutral components onto the SM gauge bosons is not strong enough to generate the correct relic abundance and avoid the overclosure of the Universe.
However, in our GUT model the correct relic density can be obtained, even in the standard thermal scenario,  through resonant annihilation mediated by the $s_6$ scalar particle, see Appendix~\ref{app:scap}. The mediator mass is mainly governed by the scalar potential quartic coupling $z_6$, which can be adjusted to get $m_{s_6}\approx 4\tev$. A small value of the trilinear coupling, at $w_{du6}\approx 100\gev$, suffices to 
produce the correct relic density exclusively through this annihilation channel. Note that, as long as $v_{s_{6}}\ll v_{s_{21}}$, this mechanism is broadly independent of $v_{s_{21}}$, which determines the mass splitting in \refeq{eq:mass_sp} and dominates the $Z'$ mass in \refeq{eq:mz}. The prediction for the $Z'$ boson mass changes minimally. We show in \reffig{fig:DD_inelastic} as a solid red line the $Z'$ mass prediction for $m_{\textrm{DM}}=2020\gev$. 
As we increase the DM mass past $\sim 2\tev$, the size of the vevs $v_{s_6}$ and $v_{s_{21}}$ become comparable. Solutions with the correct relic density can easily be found with significantly larger $m_{\textrm{DM}}$ but the mass of the $Z'$ gauge boson will increase, given equivalent $\delta m_{\textrm{DM}}$, in correlation with $m_{\textrm{DM}}$.  
\item We have discussed in \refsec{sec:dm_pheno}
that boundary conditions stemming from fixed point~(\ref{eq:fp}) 
lead to a scenario with two-component DM. An SU(2)-doublet WIMP with $m_{\textrm{DM}}< 1\tev$ annihilates too efficiently in the early Universe, thus 
providing a subdominant component of the relic abundance. On the other hand, an additional component can easily be provided in our model by the resonant annihilation of Majorana fermion~$N_3$ in \refeq{eq:physmas}, with the exchange of a heavy scalar mediator with mass $m_{s_{21}}\approx 2 m_{N_3}$, which can result from the adjustment of the parameters of the scalar potential (see  Appendix~\ref{app:scap}). A suppression in the local density of the pseudo-Dirac DM particle~$N_1$ will suppress the corresponding solar capture rate
for equivalent mass splitting $\delta m_{\textrm{DM}}$. On the other hand, while in presence of a local-density suppression a pseudo-Dirac DM with $\delta m_{\textrm{DM}}\approx 500\kev$ can narrowly evade the capture-rate bounds from the U and Pb elements in the Sun, which are fairly flat in $\delta m_{\textrm{DM}}$\cite{Pospelov:2026ewn}, 
the situation can hardly improve any further, due to the very steep nature of the Fe capture rate. 
Overall, we expect a slight weakening of the lower bound on the mass splitting, to approximately 
$\delta m_{\textrm{DM}}\gsim 500\kev$\cite{Pospelov:2026ewn}. 
Note that, for two-component DM, DM masses lighter than $\sim 1\tev$ all 
correspond to the dashed red line featured in \reffig{fig:DD_inelastic} because, while $m_{N_1}$ is dominated by vev $v_{s_6}$, $\delta m_{\textrm{DM}}$ and $m_{Z'}$ are dominated by $v_{s_{21}}$. One can thus see that $\delta m_{\textrm{DM}}\gsim 500\kev$ allows one to narrowly evade the solar capture bound under the
SHM+LMC assumption, without interfering with our prediction of a $Z'$ gauge boson with $m_{Z'}\approx 5.5\tev$. 
\end{enumerate}

%%%%%%%%%%%%%%%%%%%%%%%%%%%%%%%%
\begin{table}[t]
\centering
\begin{tabular}{|c|c|c|c|c|c|c|c|c|c|c|}
\hline
 & $v_{s_6}$ & $v_{s_{21}}$ & $m_{N_1}$ & $m_{N_3}$ & $m_{s_6}$ & $m_{s_{21}}$ & $z_6$ & $z_{21}$ & $w_{du6}$ & $w_{2166}$ \\
 \hline
BP1 & 5.20 & 8.00 & 2.02 & \textcolor{gray}{5.77} & 4.15 & \textcolor{gray}{6.74} & 0.316 & \textcolor{gray}{0.35} & 0.1 & \textcolor{gray}{-0.15} \\
\hline
BP2 & 1.60 & 7.40 & 0.62 & 5.34 & \textcolor{gray}{1.63} & 10.90 & \textcolor{gray}{0.6} & 1.08 & \textcolor{gray}{0.3} & -0.95 \\
\hline
\end{tabular}
\caption{The parameters of scalar potential~(\ref{eq:lowpot}) that are required to obtain the correct relic density in the two scenarios described in \refsec{sec:dm_solar} are shown in black. All dimensionful quantities are given in\tev. }
\label{tab:relic}
\end{table}
%%%%%%%%%%%%%%%%%%%%%%%%%%%%%%

To check the consistency of the two scenarios described above we have implemented the model in \texttt{SARAH}\cite{Staub:2013tta,Staub:2012pb,Staub:2010jh,Staub:2009bi}, computed the particle spectra using \texttt{FlexibleSUSY}\cite{Athron:2014yba,Athron:2017fvs}, checked the viability of the Higgs sector\cite{Athron:2021kve,Kotlarski:2026zmq}, and computed the DM relic density using \texttt{micrOMEGAs}\cite{Belanger:2026asz}.
In \reftable{tab:relic} we present the parameter values required to obtain the correct relic density, $\Omega h^2\approx 0.12$, for two benchmark points associated, respectively, with each scenario. The parameters given in gray do not contribute to the relic abundance calculation for the case in question.

%%%%%%%%%%%%%%%%%%%%%%%%%%%%%%%%%%%%%%%%%%%%%%%%%%%%%%%%%%%%%
\section{Collider signatures\label{sec:collider}}
%%%%%%%%%%%%%%%%%%%%%%%%%%%%%%%%%%%%%%%%%%%%%%%%%%%%%%%%%%%%%%

%%%%%%%%%%%%%%%%%%%%%%%%%%%%%%%%%%%%%%%%%%%%%%%%%%%%%%%%%%%%%%%%%%%%%%%
 \begin{figure}[t]
	\centering%
		\includegraphics[width=0.4\textwidth]{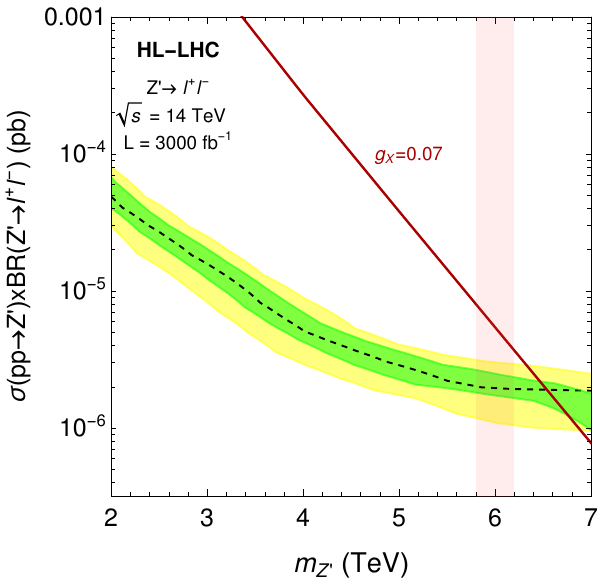}
\caption{Leading order cross section for $pp \to Z^\prime\to l^+ l^-$, with $l = e, \mu$, for $\sqrt{s}=14\tev$~(red solid) at HL-LHC. The predicted $Z'$ mass range corresponding to the blue band in \reffig{fig:DD_inelastic} is shown as a pink vertical band. Projected 95\% and 68\%~intervals for $5\sigma$ discovery are taken from Ref.\cite{CMS-PAS-FTR-21-005}.
}
\label{fig:hllhc2}
\end{figure}
%%%%%%%%%%%%%%%%%%%%%%%%%%%%%%%%%%%%%%%%%%%%%%%%%%%%%%%%%%%%%%%%%%%%%

The predicted $Z'$ mass in the SHM+LMC case lies at $\sim 6\tev$. It is thus very likely to be observed already at the HL-LHC. 
The expected exclusion reach for heavy $s$-channel resonances in the mass range 2--7\tev\ was derived in Ref.\cite{CMS-PAS-FTR-21-005}.  In the most 
sensitive final-state channel, $l^+ l^-$, one expects limits on the corresponding cross section of around $10^{-5}$-- $10^{-6}\,\textrm{pb}$ (see Fig.~3 of Ref.\cite{CMS-PAS-FTR-21-005}).

To quantify the constraining power of the HL-LHC in our model, we focus on a simplified scenario containing only a $Z'$ boson, 
coupled to the SM fermions with coupling strength $g_X$ and gauge charges as given in \reftable{tab:q_numbers}.
The total decay width of $Z'$, for $m_{Z'} \gg M_{Z}$, reads\cite{Chikkaballi:2025pnw}
\begin{equation}\label{eq:width}
\Gamma_{Z'}= \frac{135}{6} \frac{g_X^2}{4\pi} m_{Z'}\,.
\end{equation}
For values of $g_X \ll 1$, as predicted by the GUT setup considered in this study, \refeq{eq:width} implies a very narrow resonance.
The relative branching ratios to different SM fermions are determined purely by the U(1)$_X$ gauge charges.
The $Z'$ therefore decays, in order of the decreasing branching ratio, into up-quarks, down-quarks, charged leptons, and neutrinos.
The decays are flavor universal for the considered range of $Z'$ masses.

In \reffig{fig:hllhc2} we show in red solid the cross section for $pp \to Z^\prime\to l^+ l^-$, with $l = e, \mu$, for the GUT model described in \refsec{sec:model} at the HL-LHC. Cross section was computed at the leading order using \texttt{MadGraph5\_aMC@NLO}\cite{Alwall:2014hca} based on \texttt{UFO}\cite{Degrande:2011ua,Darme:2023jdn} model generated by \texttt{SARAH}\cite{Staub:2013tta,Staub:2012pb,Staub:2010jh,Staub:2009bi}. Yellow and green shaded bands indicate, respectively, projected 68\% and 95\%  intervals for the $5\sigma$ discovery potential taken from Ref.\cite{CMS-PAS-FTR-21-005}. For the value of $g_X(\mu=1\tev) = 0.07$ predicted by the GUT boundary conditions, a $Z'$ mass in the range 5.5--6.5\tev\ (pink vertical band), as dictated by the size of the DM mass splitting that fits the LZ event in the SHM+LMC case (cf.~\reffig{fig:DD_inelastic}), falls within the projected reach, offering promising prospects of future discovery. 

%%%%%%%%%%%%%%%%%%%%%%%%%%%%%%%%%%%%%%%%%%%%%%%%%%%%%%%%%%%%%%%%%%%%%%%
 \begin{figure}[t]
	\centering%
		\includegraphics[width=0.4\textwidth]{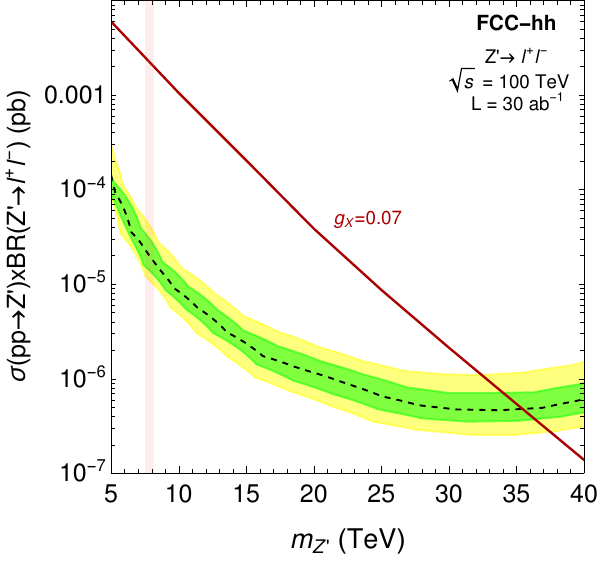}
\caption{Leading order cross section for $pp \to Z^\prime\to l^+ l^-$, with $l = e, \mu$, for $\sqrt{s}=100\tev$~(red solid) at FCC-hh. The predicted $Z'$ mass range corresponding to the green band in \reffig{fig:DD_inelastic} is shown as a pink vertical band. Projected 95\% and 68\%~CL exclusion limits are taken from Ref.\cite{Helsens:2019bfw}.
}
\label{fig:fcc2}
\end{figure}
%%%%%%%%%%%%%%%%%%%%%%%%%%%%%%%%%%%%%%%%%%%%%%%%%%%%%%%%%%%%%%%%%%%%%

As can be seen in~\reffig{fig:DD_inelastic}, the $Z'$ gauge boson associated with the SHM only case lies at larger mass, $\sim 8\tev$, which is outside of the reach of the HL-LHC. For the sake of completeness, we have calculated the potential for observation of such a resonance at the FCC$-hh$. The expected exclusion reach of the FCC for heavy $s$-channel resonances in the mass range 10--50\tev\ was derived in Ref.\cite{Helsens:2019bfw}. In the most 
sensitive final-state channel, $l^+ l^-$, one expects limits on the corresponding cross section around $10^{-6}$-- $10^{-7}\,\textrm{pb}$ (see Fig.~2 of Ref.\cite{Helsens:2019bfw}; cf.~also Fig.~2 of Ref.\cite{Brade:2025dlk}). In \reffig{fig:fcc2} we show in red solid the cross section for $pp \to Z^\prime\to l^+ l^-$, with $l = e, \mu$, assuming $\sqrt{s}=100\tev$ at the FCC-$hh$. Yellow and green shaded bands indicate, respectively, projected 68\% and 95\% CL exclusion limits taken from Ref.\cite{Helsens:2019bfw}. For $g_X(\mu=1\tev) = 0.07$, $m_{Z'}\approx 8\tev$ (pink vertical band) falls squarely within the reach of FCC-$hh$. Our result is in broad agreement with Fig.~1.1 of Ref.\cite{FCC:2018vvp}, which estimates 
that the potential discovery in the dilepton channel of a $Z'$ gauge boson in the Sequential~SM
would require only $\sim 1\textrm{fb}^{-1}$ of data at FCC-$hh$. 

\begin{figure}
  \centering
  \vspace{-0.4cm}
  \includegraphics[width=0.4\textwidth]{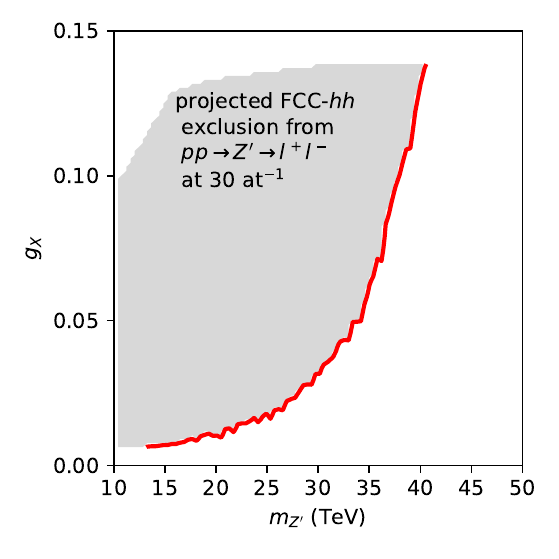}
  \vspace{-0.1cm}
  \caption{Projected 95\% CL exclusion limit (shown as red solid line) on $g_X$ versus~$m_{Z'}$, assuming $\sqrt{s}=100\tev$ and 30~ab$^{-1}$ of integrated luminosity at FCC-$hh$.
  The limit is extracted using the projected $pp \to Z' \to l^+l^-$ exclusion cross section provided in Ref.\cite{Helsens:2019bfw}.
  }
  \label{fig:fcc}
\end{figure}

To facilitate comparison with other GUT-induced scenarios featuring a $Z'$ boson, and also to capture the variety of solutions at large $m_{Z'}$ mentioned at the end of point~1 in \refsec{sec:dm_solar}, we show 
in Fig.~\ref{fig:fcc}, in the $(m_{Z'},g_X)$ plane, the parameter space potentially excluded by the FCC-$hh$ heavy-resonance search. The total branching ratio to leptons is $\text{BR}(Z^\prime \to l^+ l^-)$ = 0.074, constant throughout the considered parameter space.
In this approach, where $m_{Z'}$ and $g_X$ are treated as free parameters, a $Z'$ boson 
with mass up to 40\tev\ could be possibly observed (depending on the value of $g_X$).

%%%%%%%%%%%%%%%%%%%%%%%%%%%%%%%%%%%%%%%%%%%%%%%%%%%%%%%%%%%%%%%%%%%%%
%%%%%%%%%%%%%%%%%%%%%%%%%%%%%%%%%%%%%%%%%%%%%%%%%%%%%%%%%%%%%%%%%%%%%

We conclude with some remarks about the prospects for heavy vectorlike charged particles in a future collider. The SU(2) doublet nature of the DM candidate implies the presence of a vectorlike tau lepton at the TeV~scale, see \refeq{eq:VL}. To what extent the vectorlike tau mixes with the SM tau is a model-dependent question and we do not attempt to provide an answer here (see, for example, Ref.\cite{CarcamoHernandez:2023wzf} for a discussion of possible LHC signatures). We just point out that even small levels of mixing could open the door to the direct discovery of such a particle in the multilepton or lepton+jet channels of a future hadron collider. As can be seen in \refeq{eq:VL}, similar considerations apply to a vectorlike bottom quark with approximately twice the mass of the DM, which is a singlet of SU(2)$_L$. The mixing with the SM bottom will determine its decay chains, but it is safe to assume that, given its relatively light mass, it may induce large excesses in the multijet channels of a 100\tev\ hadron collider. 

Finally, the need to avoid the solar capture bound has lead us to consider the scenarios discussed in \refsec{sec:dm_solar}, which need the presence of heavy (pseudo-)scalar fields if one wants to accommodate the correct relic density within the traditional freeze-out framework. A rich spectrum is likely to contribute a multiplicity of extra signals at future colliders. As the specific decay channels and strengths of these particles are very model-dependent, we leave a detailed analysis of those signatures for future work.  

%%%%%%%%%%%%%%%%%%%%%%%%%%%%%%%%%%%%%%%%%%%%%%%%%%%%%%%%%%%%%%%%%%%%%%%%%%%%%%%%
\section{Conclusions\label{sec:conc}}
%%%%%%%%%%%%%%%%%%%%%%%%%%%%%%%%%%%%%%%%%%%%%%%%%%%%%%%%%%%%%%%%%%%%%%%%%%%%%%%%

In this paper, we have investigated future-collider signatures indirectly 
associated with the recently observed recoil event at LUX-ZEPLIN. A popular explanation for the origin of the LZ event 
is the possible inelastic scattering of a Higgsino or Higgsino-like particle, where the two neutralino states are separated
by approximately $\delta m\approx 350\kev$, if the Galactic velocity distribution follows the SHM, or $\delta m\approx 500\kev$, if the LMC boost is added to it (recently established bounds from the solar capture of DM in the Sun strongly favor the latter solution). In the MSSM the observed signal implies a split sparticle spectrum, with the particles that indirectly determine the size of $\delta m$, the gauginos, too heavy for one to realistically expect their direct observation at any future collider. Here we focus on an alternative to the Higgsino, based on non-supersymmetric Grand Unified Theory. The DM candidate has very similar properties to the Higgsino's but the splitting of neutral states is associated, via the GUT construction, with the mass of new gauge bosons much lighter than the corresponding MSSM gauginos. 
Thus, as a direct consequence of the LZ event, one expects for GUT-induced models promising prospects for discovery at future colliders. Excitingly, such signals are very likely to show up already at the HL-LHC, and can then be confirmed with gargantuan evidence in the very early run of an FCC-$hh$ detector.

We have explored this scenario in a specific example. Our model descends from a SU(6) GUT and it predicts a new $Z'$ gauge boson with a relatively narrow width, mass around 6-8\tev, and couplings of SM strength to the SM particles. We have numerically computed the cross section for a signal in the SM dilepton channel at a $pp$ collider with $\sqrt{s}=14\tev$ and $\sqrt{s}=100\tev$ c.o.m.~energy. The signal falls squarely within the published projected sensitivities of the HL-LHC and FCC-$hh$ to heavy $Z'$ resonances. Additional collider signatures of the model involve the early-run direct observation of heavy vector-like leptons and quarks, and heavy scalars. 

Our study shows that the pseudo-Dirac DM explanation of the LZ event, if it survives further scrutiny, has the right properties to open the door to a wealth of exciting new discoveries in the planned accelerators of the future.     

%%%%%%%%%%%%%%%%%%%%%%%%%%%%%%%%%%%%%%%%%%%%%%%%%%%%%%%%%%%
\begin{acknowledgments}
WK is supported by the National Science Centre (Poland) grant 2022/\allowbreak47/\allowbreak D/\allowbreak ST2/\allowbreak03087. EMS is supported in part by the National Science Centre (Poland) under the research Grant No.~2020/38/E/ST2/00126.
\end{acknowledgments}

%%%%%%%%%%%%%%%%%%%%%%%%%%%%%%%%%%%%%%%%%%%%%%%%%%%%%%%%%%%
\appendix
%\section*{Appendices}
\addcontentsline{toc}{section}{Appendices}

%%%%%%%%%%%%%%%%%%%%%%%%%%%%%%%%%%%%%%%%%%%%%%
\section{Low-scale scalar potential}
\label{app:scap}

We recall in this appendix the scalar potential of the model introduced in Ref.\cite{Chikkaballi:2025pnw}. Analytic approximations for the scalar masses can be found in Appendix~A.2 of that work.

After the SU(6) and SU(5) GUT symmetries are broken, we are left with the following potential for the SU(2)$_L$ doublets $H_u$, $H_d$ and the singlets $s_6$, $s_{21}$:
\begin{multline}\label{eq:lowpot}
V_{\textrm{low}}= \mu_d^2\, H_d H_d^{\dag}+\mu_u^2\, H_u^\dagger H_u+\mu_6^2|s_6|^2+\mu_{21}^2 |s_{21}|^2\\
+\,z_d\left(H_d H_d^{\dag}\right)^2+z_u\left(H_u^\dagger H_u\right)^2+z_6 |s_6|^4\\
+\,z_{21} |s_{21}|^4
+z_{d6}H_d H_d^{\dag} |s_6|^2+z_{d21} H_d H_d^{\dag} |s_{21}|^2\\
+\,z_{u6} H_u^\dagger H_u |s_6|^2+z_{u21}H_u^\dagger H_u |s_{21}|^2
+z_{du}\left(H_d H_d^{\dag}\right)\left( H_u^\dagger H_u\right)\\
+\,z_{dudu}H_d H_u\left(H_d H_u\right)^{\dag}+z_{621}|s_6|^2 |s_{21}|^2\\
+\left(z_{ud621} H_d H_u s_6^{\dag} s_{21}^{\dag}+w_{du6}H_d H_u s_6\right.\\
\left.+\,w_{2166}\, s_{21} s_6^2+\textrm{H.c.}\right),
\end{multline}
where the $z_i$ indicate generic quartic couplings and the $w_j$ generic trilinear couplings. 

The couplings of the scalar potential~(\ref{eq:lowpot}) become important when one tries to evade the solar-capture bounds on the inelastic DM mass splitting. As we discuss in \refsec{sec:dm_solar}, in fact, the presence of a resonance with a (pseudo-)scalar mediator can avoid the overabundance of DM for masses of the WIMP well above~1\tev.  As we show in \reftable{tab:relic}, in the one-component DM scenario the only parameters that are relevant to the relic density computation are $z_6$, which determines the mediator mass, and $w_{du6}$, which determines the coupling strength of the mediator to the Higgs boson. When considering instead a two-component scenario, one will need a resonance with a heavier scalar particle, whose mass and interaction strength are regulated, respectively, by $z_{21}$ and $w_{2166}$. Note that we work under the assumption of small mixing between the heavy and light scalar particles, which implies $z_{621},z_{ud612}\ll 1$.  

\bigskip

\bibliography{mybib}

\end{document}